\documentclass[aps,showpacs,prx,amssymb,amsmath,twocolumn]{revtex4-2}
\usepackage{graphicx}
\usepackage{amssymb}
\usepackage{amsmath}
\usepackage{amsthm}
\usepackage{bm}
\usepackage{xcolor} 
\usepackage{array}
\usepackage{multirow}
\usepackage{tabularx}
\usepackage{booktabs}
\usepackage{textcomp}
\usepackage{mathtools}
\usepackage{hyperref}

\usepackage{qcircuit}

\usepackage{bbm}
\usepackage{cancel}
\usepackage{comment}
\usepackage{physics}

\begin{document}

\title{Optical tweezers throw and catch single atoms}

\author{Hansub Hwang$^1$, Andrew Byun$^1$, Juyoung Park$^1$, Sylvain de Léséleuc$^{2,3}$, and Jaewook Ahn$^{1}$}
\email{*jwahn@kaist.ac.kr}
\address{$^1$Department of Physics, KAIST, Daejeon 34141, Korea}
\address{$^2$Institute for Molecular Science, National Institutes of Natural Sciences, Okazaki 444-8585, Japan}
\address{$^3$SOKENDAI (The Graduate University for Advanced Studies), Okazaki 444-8585, Japan}

\date{\today}

\begin{abstract} \noindent
Single atoms movable from one place to another would enable a flying quantum memory that can be used for quantum communication and quantum computing at the same time. Guided atoms, e.g., by optical tweezers, provide a partial solution, but the benefit of flying qubits could be lost if they still interact with the guiding means. Here we propose and experimentally demonstrate freely-flying atoms that are not guided but are instead thrown and caught by optical tweezers. In experiments, cold atoms at 40~$\mu$K temperature are thrown up to a free-flying speed of 0.65 m/s over a travel distance of 12.6~$\mu$m at a transportation efficiency of 94(3)\%, even in the presence of other optical tweezers or atoms en route. This performance is not fundamentally limited but by current settings of optical tweezers with limited potential depth and width. We provide a set of proof-of-principle flying atom demonstrations, which include atom transport through optical tweezers, atom arrangements by flying atoms, and atom scattering off optical tweezers. Our study suggests possible applications of flying atoms, not only in fundamental studies such as single-atom low-energy collisions, but also non-photon quantum communication and flying-qubit-based quantum computing.
\end{abstract}

\maketitle

\section{Introduction}
\noindent
Optical tweezers are a versatile tool in modern science, being widely used to trap and guide small particles such as atoms~\cite{AshkinPRL1970_tweezer, AshkinOL1986_tweezer}, molecules~\cite{JohnMDoyle_molecule}, micro-beads~\cite{SvobodaOL1994_bead}, and biological objects~\cite{AshkinOL1987_cell_bio}. Single particles manipulated at a few micrometer distance by optical tweezers are well suited to study, i.e., control and investigate, their interaction nature among themselves and with others, even at the quantum level~\cite{Rydberg_blockade_urban, Rydberg_blockade_geatan, quantum_level_interaction}. In that regards, single atoms in optical tweezers are drawing keen attention because of their promising usage as an elementary quantum information carrier. In recent years, there are as many as a few hundred single atoms dynamically rearranged with optical tweezers for defect-free atom arrays~\cite{EndresSci2016_1daod, BarredoSci2016_2daod, KimNatComm2016_2dslm,LeeOE2016_3dslm}, and related Rydberg atom experiments are making a tangible progress in quantum computing and quantum simulations~\cite{BernienNat2017_1DIsing, KimPRL2018_1DIsing, GuardadoSanchezPRX2018_2DIsing, LienhardPRX2018_2DIsing, KimPRXq2020_3DIsing, BarredoPRL2015_1D2DXY, SchollPRXq2022_1D2DXXZ}. 

Optical tweezers are a useful tool, not only to trap atoms statically but also to dynamically guide them through desirable paths. However, these paths are to avoid optical tweezer collisions, especially in a crowded atom array, which often result in unwanted atom loss. Being inspired by a recent experiment of coherent atom transportation amid quantum operations~\cite{BluvsteinNature2022}, we consider a way to deliver an atom between places without using an optical tweezer in between, i.e., to use the optical tweezer as an atom accelerator (i.e., an atom thrower) and decelerator (i.e., an atom catcher), but not as an atom carrier.

The physics of the optical tweezer trapping of an atom is well understood as a classical particle dynamics in a truncated harmonic potential $U(\xi)$ given by
\begin{equation} \label{U}
U(\xi) = \frac{U_0}{d^2}(\xi-d)(d+\xi)
\end{equation}
where $U_0$ and $d$ are the potential depth and width of the optical tweezer, and $\xi$ is the atom displacement from the center of the optical tweezer. An atom of an energy higher than $U_0$ or displaced more than $d$ escapes from the optical tweezer. Here we assume the atom does not manifest quantum features as the temperature is not comparable to the energy of the vibration quanta~\cite{HickmanPRA2020, LamPRX2021}. Then, the equation of motion of the atom is given by
\begin{equation}
\xi(t) =-\frac{\ddot{x}}{\omega^2} + A \cos(\omega t + \phi)
\label{eq_sol}
\end{equation}
in terms of the atom displacement $\xi$ with respect to the position $x$ of the optical tweezer, where $\omega=\sqrt{2U_0/md^2}$ is the trap frequency, $A=\sqrt{(\xi_0-{\ddot{x}}/{\omega^2})^2+({\dot{\xi}_0}/{\omega})^2}$ is the oscillation amplitude, and $\phi=-\tan^{-1} [{\omega\dot{\xi}_0}/(\omega^2\xi_0-{\ddot{x}})]$ is the phase, given that the initial displacement and velocity are $\xi_0$ and $\dot{\xi}_0$, respectively.

In an accelerating optical tweezer of $\ddot{x}=a>0$, the atom in the moving frame of the optical tweezer oscillates back and forth about the equilibrium point $\langle \xi \rangle=-a/\omega_2$. As the atom on average lags behind the optical tweezer, if the maximum negative displacement exceeds the width of the optical potential, the optical tweezer loses the atom behind, which defines the maximal acceleration, $a_{\rm max}=U_0/md$, of an atom in an accelerating optical tweezer. In order to use the optical tweezer as an atom decelerator, we first make an freely-flying atom by accelerating the atom and after a certain distance of the atom's free flying we recapture and then decelerate the atom, e.g., with $\ddot x=-a$. Then the available range of $a$ is determined by both the throwing and catching conditions. While the successful throwing is mostly conditioned by $a\le a_{\rm max}$, as above, the successful catching is not as simple as the throwing, which is to be investigated below.

\section{Theoretical analysis}  
\begin{figure*}[htbp]
  \centering
\includegraphics[width=1.0\textwidth]{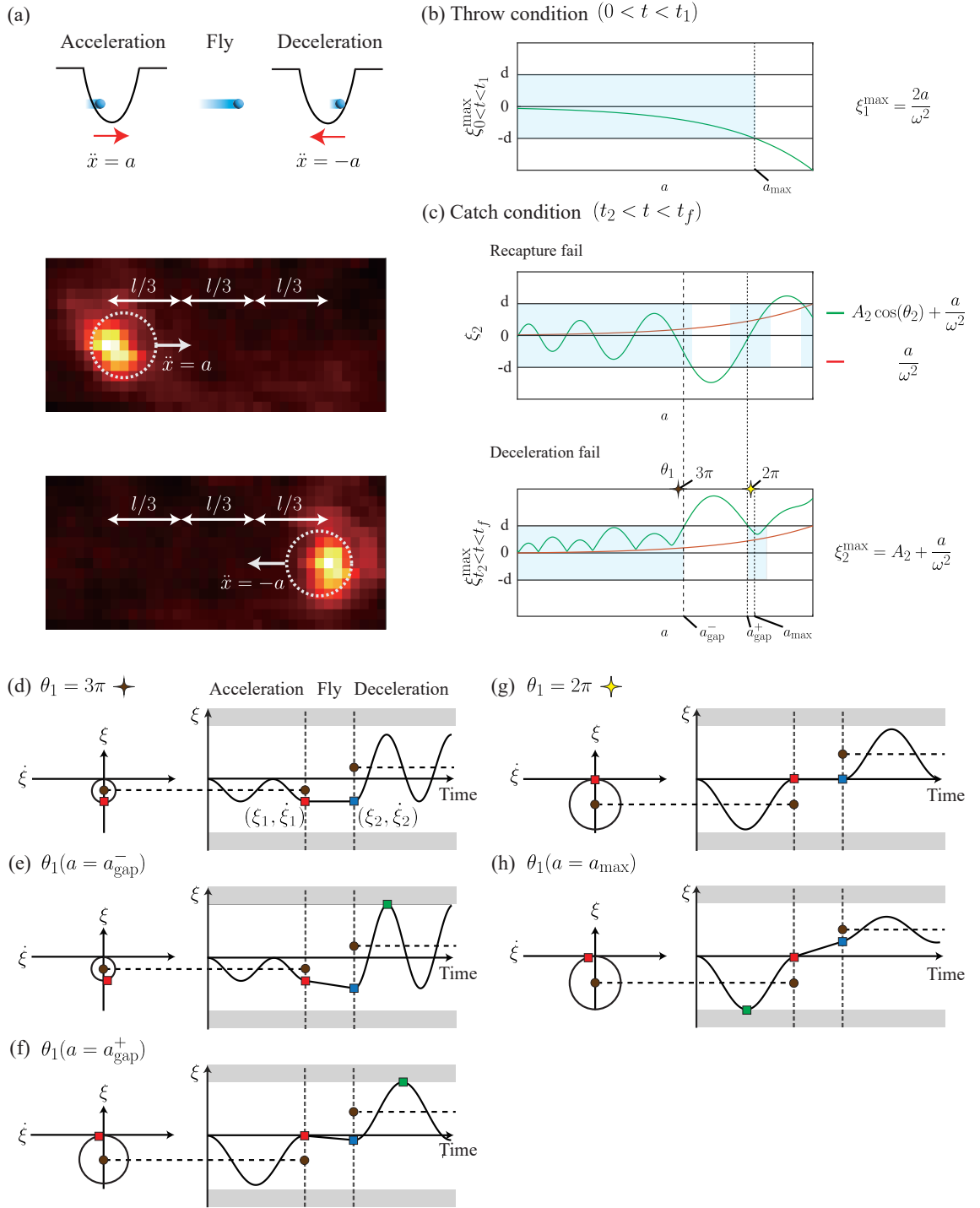}
\caption{Atom throw-and-catch by an optical tweezer.  (a) An atom is thrown by an accelerating optical tweezer of $\ddot{x}=a$, released for a free flight, and then caught by an decelerating optical tweezer of $\ddot{x}=-a$. (b) An atom image in the acceleration stage. (c) An atom image in the decelerating stage.  (d) A successful throw is determined by the maximal displacement $\xi_{\rm max}(0<t<t_1)$ within $-d$ and $d$, requiring $a<a_{\rm max}$. (e) A successful catch is determined by the recapture ($-d<\xi_2<d$) and deceleration ($-d<\xi_{\rm max}(t>t_2)<d$) conditions. (f-i) Atom dynamics in the $\xi$-$\dot{\xi}$ diagram (left) and $\xi(t)$ trajectories (right) for various release phases $\theta_1=\theta(t=t_1)$: (f) $\theta_1=3\pi$. (g) $\theta_1(a=a^-_{\rm gap})$, (h) $\theta_1(a=a^+_{\rm gap})$, (i) $\theta_1=2\pi$, and (j) $\theta_1(a=a_{\rm max})$, where the black circles indicate the equilibrium points, red and blue squares the displacements respectively at the release ($t=t_1$) and recapture ($t=t_2$), and the green squares the atom escapes (i.e., $\xi=d$ or $-d$).}
\label{Fig1}
\end{figure*}

\noindent
We consider an optical tweezer holding an atom is accelerated as in Fig.~\ref{Fig1}(a). The optical tweezer is first accelerated with $\ddot{x}=a$ along the $x$ direction for a distance $x_1$,  then  turned off for the atom to freely fly till $x_2$, and turned back on and decelerated with $\ddot{x}=-a$ till a complete stop at $x_f=l$, i.e.,
\begin{eqnarray}
\ddot{x} =
\begin{cases}
a & \mbox{for} \quad 0<t<t_1 \quad\mbox{(acceleration)}  \\
0 & \mbox{for} \quad t_1<t<t_2 \quad\mbox{(free-flying)}  \\
-a & \mbox{for} \quad t_2<t<t_f \quad\mbox{(deceleration)}
\end{cases}
\label{xddot}
\end{eqnarray}
where $t_1$ and $t_2$ denote the times when the optical tweezer is located at $x_1$ and $x_2$, respectively. Snap shot images of an as-traveling atom are shown in Figs.~\ref{Fig1}(a) and (c) during the acceleration and deceleration stages, respectively. For the sake of convenience, we choose equal distances of acceleration, free-flying, and deceleration (i.e., $x_1=l/3$ and $x_2=2l/3$), in which the characteristic times are given by $t_1=\sqrt{2l/3a}$, $t_2=\sqrt{3l/2a}$, and $t_f=\sqrt{25l/6a}$. 

The solution of Eqs.~\eqref{eq_sol} and \eqref{xddot} for an initially stationary atom (i.e, $\xi_0=\dot{\xi}_0=0$) is given by 
\begin{eqnarray}
\xi (t) =
\begin{cases}
-\frac{a}{\omega^2}+A_1 \cos{(\omega t)} & \mbox{for} \quad 0<t<t_1   \\
\xi_1 + \dot{\xi}_1 (t-t_1)  & \mbox{for} \quad t_1<t<t_2  \\
\frac{a}{\omega^2} + A_2 \cos{\left((\omega (t-t_2)+\theta_2\right)} & \mbox{for} \quad t_2<t<t_3
\end{cases}
\label{eq_sol2}
\end{eqnarray}
where $A_1={a}/{\omega^2}$, $A_2=\sqrt{(\xi_2-{a}/{\omega^2})^2+({\dot{\xi}_2}/{\omega}})^2$ are the oscillation amplitudes during the acceleration and deceleration, respectively, $\xi_{1}= \frac{a}{\omega^2}(\cos \theta_1-1)$, $\xi_2=\frac{a}{\omega^2}+A_2\cos\theta_2$, $\dot{\xi}_{1}=\dot{\xi}_2=-\frac{a}{\omega}\sin\theta_1$, $\theta_1=\omega t_1$, and $\theta_2=\sin^{-1}\left(\frac{a}{\omega^2 A_2}\sin\theta_1\right)$ are the displacements, velocities, and oscillation phases, respectively, at the start ($t=t_1$) and end ($t=t_2$) of the free-flying.

For a successful throw and catch, the atom needs to be kept in the optical tweezer during the acceleration and deceleration stages, i.e., $-d\le {\xi(t)} \le d$ for $0<t<t_1$ and $t_2<t<t_f$. In that regards, there are three distinct atom escape scenarios: (1) the atom could escape at the acceleration stage; (2) the atom could fail being recaptured at $t=t_2$; (3) the atom could escape at the deceleration stage. 

(1) First, in the acceleration stage ($0<t<t_1$), the atom escape is determined by the maximal (negative) displacement during this time interval, which is given by either $|\xi|^{\rm max}(0<t<t_1)=2a/\omega^2$ or $|\xi(t_1)|$, depending on whether there are at least $1/2$ oscillations until $t_1$ (i.e., $\theta_1=\omega t_1>\pi$) or not. As $t_1$ is a function of $l$ (the total travel length), we choose $l$ to satisfy the former ($\theta_1>pi$) in our experimental consideration (i.e., $l>3\pi^2d/4$). Then, the atom escape occurs when $a>a_{\rm max}=d\omega^2/2$. In Fig.~\ref{Fig1}(d), we plot this maximal (negative) displacement, $-|\xi|^{\rm max}(0<t<t_1)=-2a/\omega^2$, as a function of $a$. Then, the atom throw condition (i.e., a safe guide until $t_1$) is given by $a<a_{\rm max}$ (the colored region). We note that the latter case of lesser than $1/2$ oscillations (i.e, $\theta_1<\pi$) could allow $a>a_{\rm max}$ but the resulting release speed is limited. 

(2) Second, at the recapture ($t=t_2$), the atom could remain outside of the optical tweezer, i.e., $|\xi_2|>d$, resulting that the optical tweezer fails to capture the atom. In Fig.~\ref{Fig1}(d) (the upper sub-figure), we plot $\xi_2$ as a function of $a$, which shows that there are gap regions between which this type of atom loss occurs. Other regions (the colored regions) allow successful atom capture. 

(3) Third, the atom could be lost during the deceleration stage ($t_2<t<t_f$), which is determined by the maximal displacement during this time interval exceeding $d$, i.e., ${\rm max}(\xi(t_2<t<t_3)=A_2+a/\omega^2>d$, where $A_2$ is also a function of $a$. In Fig.~\ref{Fig1}(e) (the lower sub-figure), we plot ${\rm max}(\xi(t_2<t<t_f))$, which shows there are a gap region of this type of atom loss and other regions allowing successful atom deceleration until $t_f$ (the colored regions). We note that, an addition to (3), there could be another case in that the atom could be lost during the acceleration but happens to fly back into the optical tweezer during deceleration. This case is analyzed to be possible for $\omega t_1 < \pi$, as otherwise such atom flies across the optical tweezer, but this is outside of our experimental region of interest. As a combined result of (1), (2), and (3), the atom throw-and-catch could be successful in two regions of acceleration, $a<a_{\rm gap}^{-}$ and $a_{\rm gap}^{+}$, where $a_{\rm gap}^{-}$ and $a_{\rm gap}^{+}$ are the accelerations satisfying the equations $A_2(a_{\rm gap}^{+})+a_{\rm gap}^{+}/\omega^2=d$ and $A_2(a_{\rm gap}^{-})+a_{\rm gap}^{-}/\omega^2=d$, respectively, as shown in Fig.~\ref{Fig1}(e).

\begin{figure*}[t]
  \centering
\includegraphics[width=0.75\textwidth]{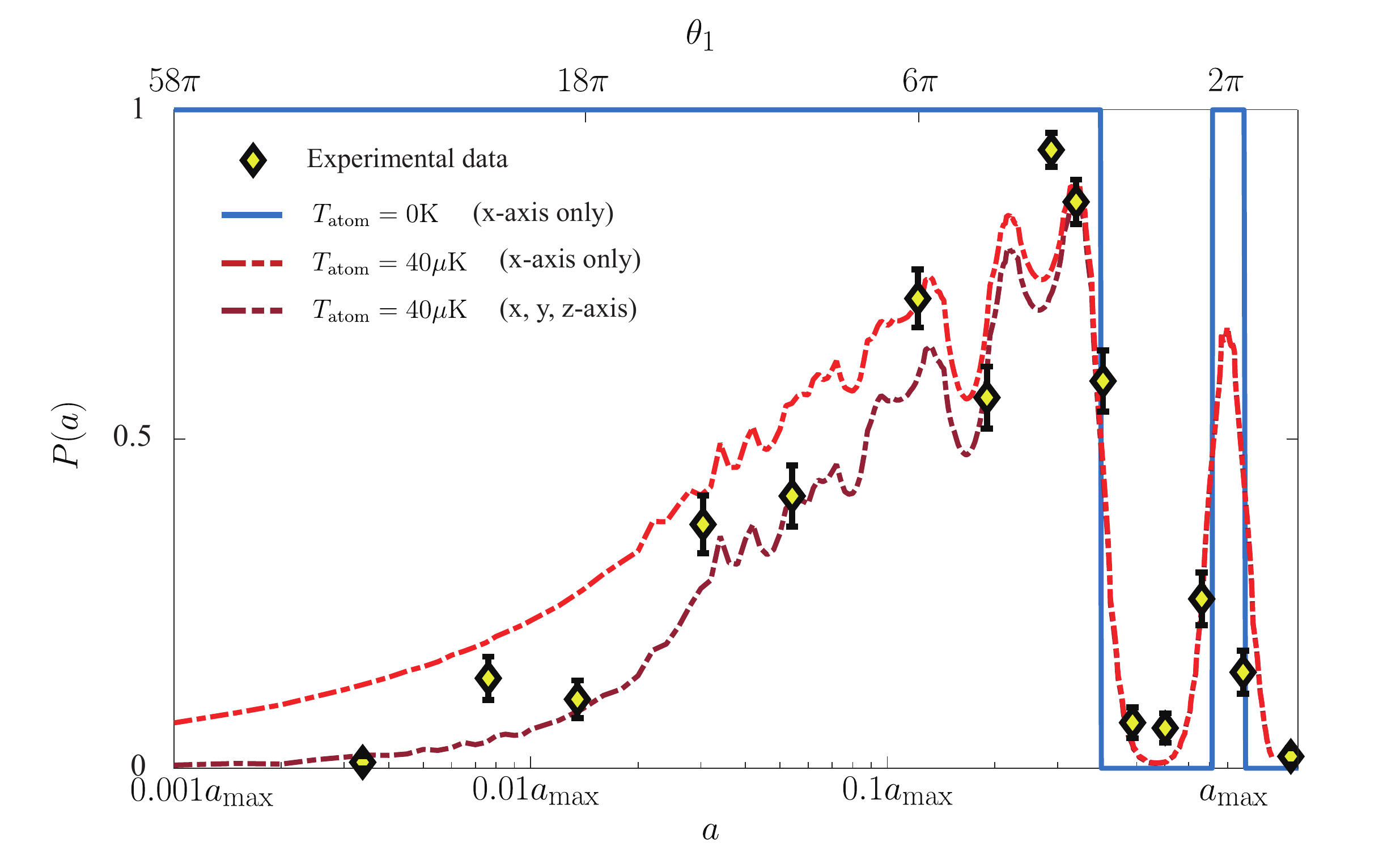}
\caption{Experimental atom throw-and-catch success probability $P(a)$ measured for various accelerations. In comparison, theory lines are shown; blue, red, and purple lines are for $T=0$~K ($x$-motion only), 40~$\mu$K ($x$ motion only), and 40~$\mu$K ($x,y,z$ motions), respectively.}
\label{Fig2}
\end{figure*}

For an intuitive understanding of the successful atom catch-and-throw, we illustrate some characteristic atom dynamics in Figs.~\ref{Fig1}(d-h), where the left and right sub-figures show the $(\xi, \dot{\xi})$ phase diagrams and $\xi(t)$ time-trajectories, respectively. In every case, the initial atom starts an oscillation about the equilibrium point $-a/\omega^2$, is then released (marked with red squares) and recaptured (blue circles), and again oscillates about the new equilibrium point $a/\omega^2$. With the increasing order of $a$, which is the decreasing order of release phase $\theta_1=\omega\sqrt{2l/3a}$, we consider $\theta_1=3\pi$, $\theta_1(a=a_{\rm gap}^-)$ that is near the former, $\theta_1(a=a_{\rm gap}^+)$, $2\pi$, and $\theta_1(a=a_{\rm max})$, where the last three are proximate to each other. First, when $\theta_1=3\pi$ as in Fig.~\ref{Fig1}(f), the atom is released with zero velocity ($\dot{\xi}(t_1)=0$) and thus recaptured with the same displacement, i.e., $\xi(t_1)=\xi(t_2)$. As a result, the recaptured atom dynamics is a simple oscillation with an amplitude three times bigger, i.e., $A_2=3A_1$. If $a$ is increased to $\theta_1(a=a_{\rm gap}^-)$, the oscillation becomes wild enough to reach the optical tweezer boundary, i.e., $\xi_2=d$, making the atom escape from the optical tweezer. The escape point is indicated with a green square in Fig.~\ref{Fig1}(g). Similarly, around $\theta_1=2\pi$, where the amplitudes before and after are the same as in Fig.~\ref{Fig1}(i), the displacement can reach either $-d$ before the release if $\theta_1(a=a_{\rm gap}^+)$, as in Fig.~\ref{Fig1}(h), or $d$ after the recapture if $\theta_1(a=a_{\rm max})$, as in Fig.~\ref{Fig1}(j). Those escape points are indicated with green squares in Fig.~\ref{Fig1}(h,j). Therefore, it can be understood that a successful atom throw-and -catch occurs either in the low acceleration zone below about $a(\theta_1=3\pi)$ or in the acceleration island zone around $a(\theta_1=2\pi)$, in both of which the atom displacement during the free flying is maintained to be small, even at high speeds, with respect to the optical tweezer.

\section{Experimental demonstration} \noindent
Experimental results of the above atom throw-and-catch scenarios are summarized in Fig.~\ref{Fig2}.
We used cold rubidium atoms ($^{87}$Rb) at 40~K. The experimental setup is similar to the one previously reported elsewhere~\cite{LeeOE2016_3dslm,BarredoSci2016_2daod,HickmanPRA2020}, except for an acousto-optic modulation add-on (AOD, DTSXY-400-820 by AA Opto electronics) which is used  to control an optical tweezer (the dynamic optical tweezer) for acceleration, flying and deceleration.  We programmed the AOD with an arbitrary waveform generator (AWG, M4i-6622-x8 by SPECTRUM instrument) of 625~MS/s sampling rate. In other experiments below which used static optical tweezers, we used a two-dimensional spatial light modulator (SLM, ODPDM-512 by Meadowlark optics) located at the Fourier plane of the atom space. The trap potential depth and width in Eq.~\eqref{U} were $U_0=1.94(15)$ or 0.76(6)~mK and $d=0.5$~$\mu$m, respectively. The atom travel length was $l=12.6$~$\mu$m, limited by the current setup. We measure the presence of the atoms using the fluorescence imaging of 5S$_{1/2}-$5P$_{3/2}$ transition~\cite{HyosubOE2019_imaging}.

In Fig.~\ref{Fig2}, measurements of the success probability $P(a)$ of the atom throw-and-catch are plotted as a function of $a$. About 120 times of the same measurements are repeated to accumulate the atom counts for each $P(a)$ of $a$ chosen from $4.20(32) \times 10^2$~m/s$^2$ to $1.85(14) \times 10^5$ ~m/s$^2$. In Fig.~\ref{Fig2}, it is observed that the probability peaks around $a(\theta_1=3\pi)=0.38 a_{\rm max}$ and is small between $a(\theta_1=3\pi)$ and $a(\theta_1=2\pi)=0.85 a_{\rm max}$, which is in a good agreement with the theoretical expectation. The details of $P(a)$ in Fig.~\ref{Fig2} are attributed to the finite temperature effect and transverse motion. The temperature effect is given by 
\begin{equation}
P(a,T)=\int P(a,E)\rho(E,T) dE,
\end{equation}
where $T$ is the atom temperature, $E=\frac{1}{2}m\dot{\xi}^2_0+\frac{1}{2}m\omega^2{\xi}^2_0$ is the initial energy, and $\rho(E,T)$ is the Maxwell-Boltzmann distribution of the atom energy in $T$~\cite{C_Tuchendler_Energy_distribution}. Numerical simulation is shown with lines in Fig.~\ref{Fig2}, where it is easy to find that this effect is dominant at high accelerations around $a(\theta_1=3\pi)$, $a(\theta_1=2\pi)$, and $a_{\rm max}$. At low accelerations, the atom escapes from the optical tweezer due to finite initial velocities, not only along the acceleration direction but also along the transverse directions. At extremely low accelerations, the success probability of atom throw and catch is approximated to
\begin{equation}
P(a,T) \simeq \sqrt{\frac{4d}{\pi l}\frac{U_0}{k_B T} }  \left(\frac{a}{a_{\rm max}}\right)^{1/2}
\label{low_a_success},
\end{equation}
where $k_B$ is the Boltzmann constant. Within the scan range of $a$ from 2.25 $\times 10^{-3} \times a_{\rm max}$ = $3.88(27) \times 10^2$~m/s$^2$ to 1.35 $\times$ $a_{\rm max}=2.33(16) \times 10^5$~m/s$^2$, which correspond to the moving times of $t_f$ ranging from 300(1)~$\mu$s to 15(1)~$\mu$s over the distance of $l$. The maximal throw-and-catch probability is measured to be $P(a_{\rm opt})=92(3)$\% (or 94(3)\% after SPAM corrections) and the optimal acceleration is $a^{\rm exp}_{\rm opt}=0.29a_{\rm max} = 5.00(35) \times 10^{4}$~m/s$^2$ being near a numerical estimation, $a_{\rm opt}=0.33 a_{\rm max}=5.69 \times 10^{4}$~m/s$^2$, obtained with the above temperature effect taken into account. Experimental errors attributed to mostly the state preparation and measurements (SPAM) were independently calibrated as $P$(zero atom| one atom)$\simeq$ 0\% and $P$(one atom|zero atom)=2\% due to atom life time in trap, which are taken into account for the rest of the data analysis.

\begin{figure}[b]
  \centering
\includegraphics[width=0.5\textwidth]{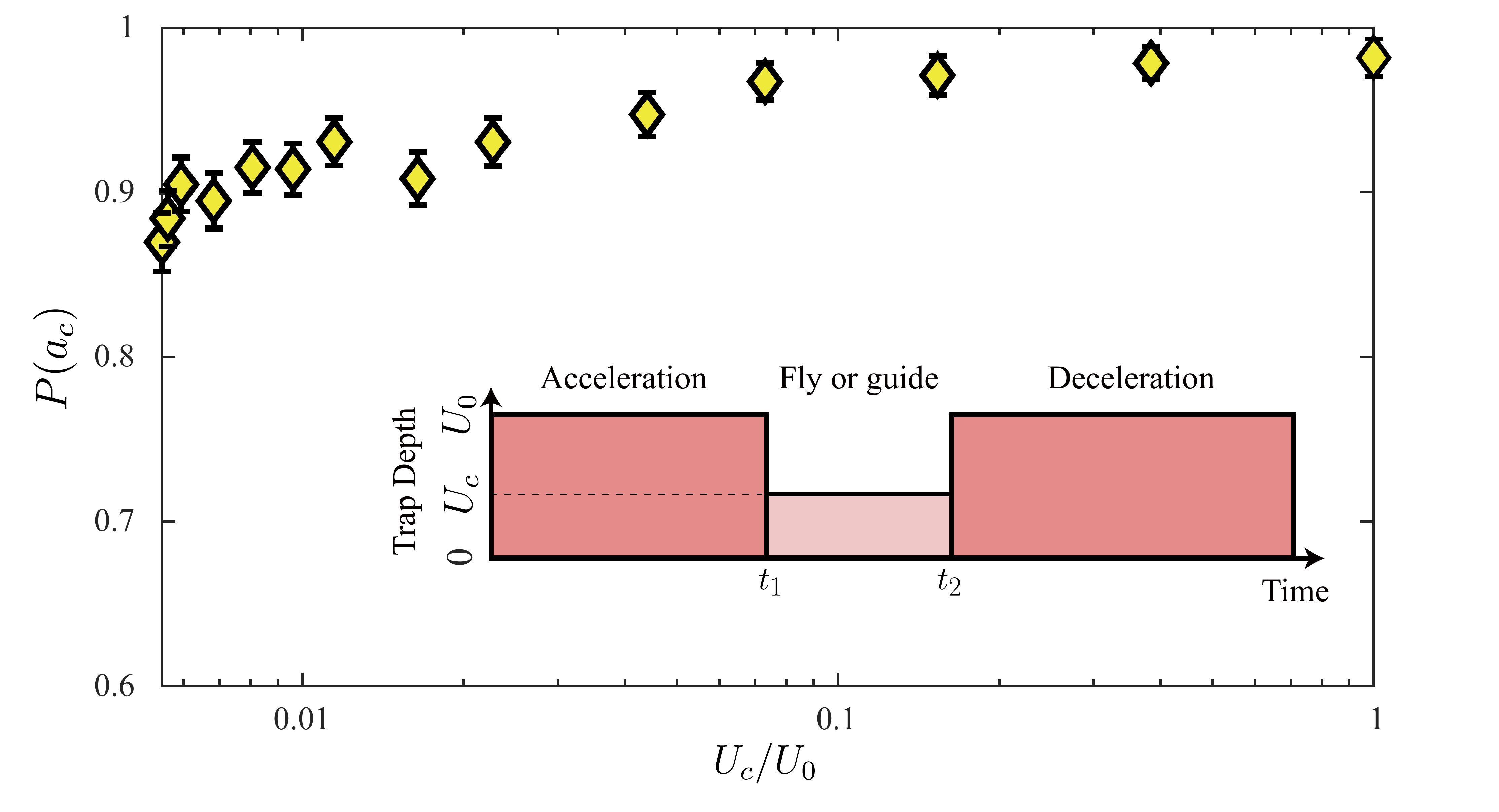}
\caption{Atom throw-and-catch success probability of a partially guided flying atom. The potential depth $U_c$ of the guiding optical tweezer during the constant speed zone ($t_1<t<t_2$) is controlled from $0$ (un-guided) to $U_0$ (fully guided).}
\label{Fig3}
\end{figure}

In the second experiment, we repeat the above measurements, but this time, without completely turning off the optical tweezer during the free-flying stage, so that we compare the throw-and-catch with the atom-guiding. The result is shown in Fig.~\ref{Fig3}, where the trap depth $U_c$ of the optical tweezer controlled during the time from $t=t_1$ and $t_2$ is changed from $U_c=0$ (the throw-and-catch) to $U_0=0.76$~mK (the atom-guiding), for a chosen $a_c$ near $a_{\rm opt}$. Data shows that the success probability $P=87(2)$\% of the throw-and-catch is not much different from $P=98(1)$\% of the atom-guiding, while there remains a gradual increase from the former to the latter due to transverse motions and experimental imperfections.

\begin{figure}[t]
  \centering
\includegraphics[width=0.5\textwidth]{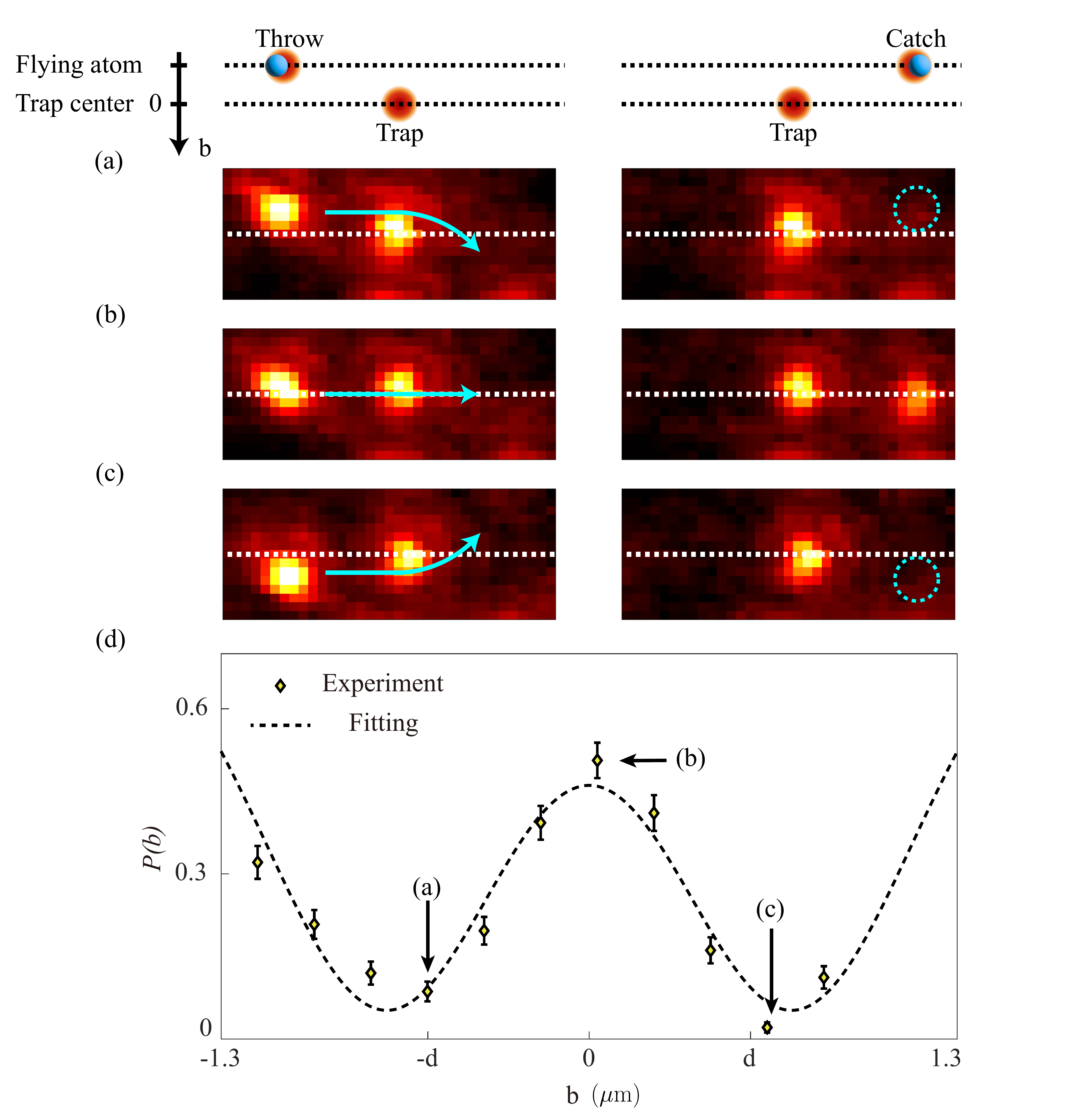}
\caption{Atom flight bent by an optical tweezer. (a-c) Atom images before and after (left and right, respectively) the collision of a flying atom and an en route optical tweezer which holds another atom; (a) Collision with positive displacement $b$, (b) $b=0$, and (c) negative $b$. (d) Measured atom throw-and-catch success probability $P(b)$ as a function of $b$, where the line is a numerical fit to $0.82-\alpha \rm{exp}(-{(b-\gamma)^2}/{\beta})-\alpha\rm{exp}(-{(b+\gamma)^2}/{\beta})$ of $\alpha=0.77$, $\beta=0.36 \mu m^2$ and $\gamma=0.72 \mu m$. }
\label{Fig4}
\end{figure}

In addition, we were able to send the flying atom through an optical tweezer. As in Fig.~\ref{Fig4}, we throw an atom so that it passes an en route optical tweezer (the static optical tweezer), which is not the one (the dynamic optical tweezer) used for the atom flying. Then the throw-and-catch probability is measured as a function of the later displacement $b$ of the static optical tweezer with respect to the atom path. For example, the atom is thrown towards the static optical tweezer sideways or straight as shown in Figs.~\ref{Fig4}(a-c) with negative $b$, $b=0$, or positive $b$, respectively.  Measured probabilities are shown in Fig.~\ref{Fig4}(d) as a function of $b$, in which the probability peaks at around $b=0$ (i.e., a straight-through pass) and is minimal at around both $b=\pm d$ (i.e., the atom path is bent by the presence of the optical tweezer). A classical analogy of this experiment is rolling a ball to a hole, which unless the path is straight through the center of the hole results in a curved path bent toward the hole. Similarly, the experimental result shows that the losses of the atom throw-and-catch is most significant around $b=\pm d$ where the significant bent occurs as expected.

\section{Discussions}  \noindent
There could be at least three practical applications of the presented method of flying atoms. First of all, flying atoms can benefit defect-free atom array formations~\cite{BarredoSci2016_2daod, LeeOE2016_3dslm}. In addition, this method of accelerating optical tweezers can be used as a non-charged particle accelerator and also as a flying quantum memory. We discuss these applications below. 

\begin{figure}[t]
  \centering
\includegraphics[width=0.5\textwidth]{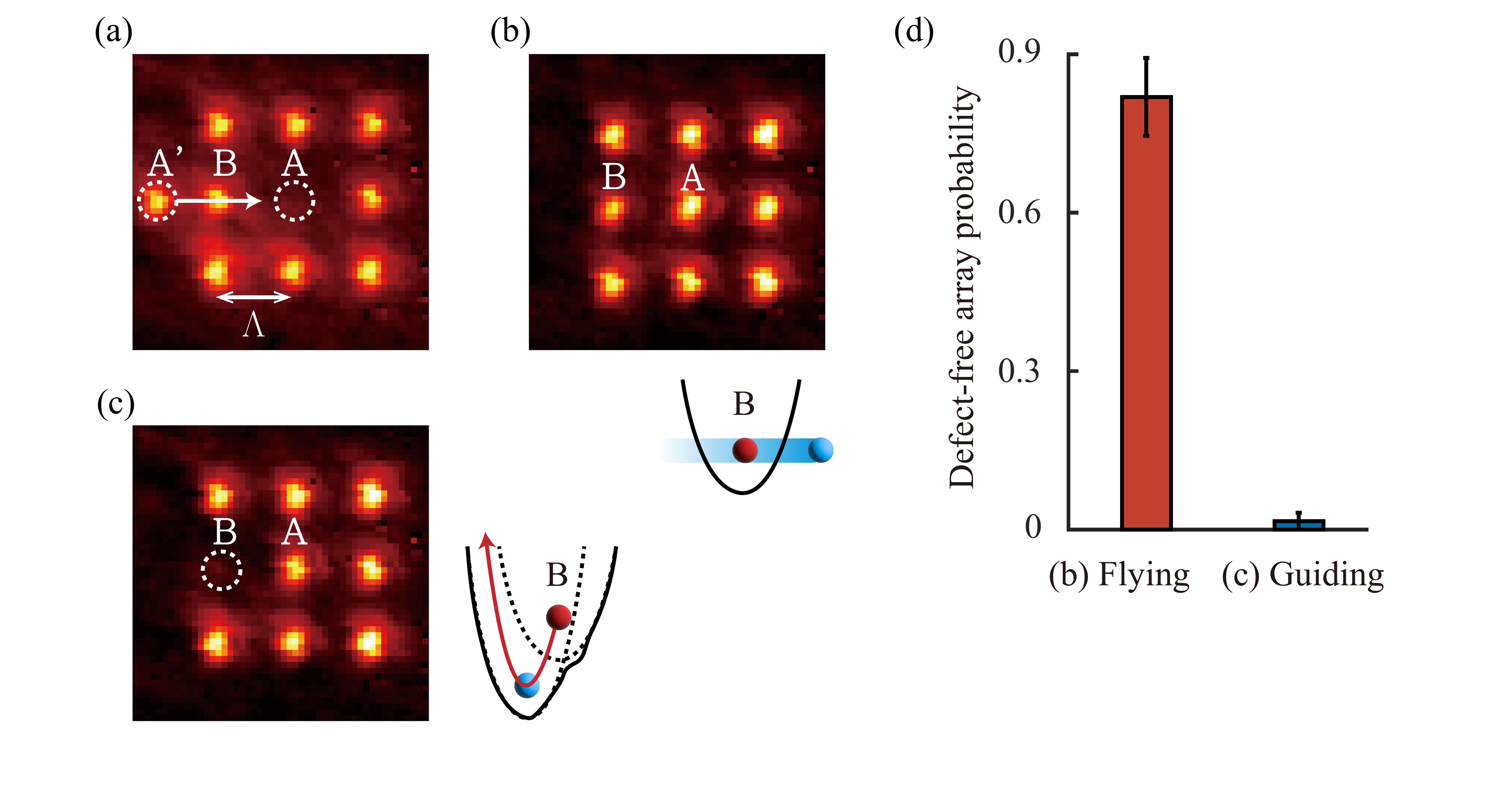}
\caption{ Defect-free array formation by a flying atom. (a) An initial defective atom array with a vacant center ($A$) and a flying atom ($A'$). (b) A resulting defect-free atom array by atom throw-and-catch from $A'$ to $A$. (c) In comparison, atom guided from $A'$ to $A$ results in a defective atom array with a vacant $B$ site. (d) Measured probabilities of defect-free atom arrays of flying and guiding atom methods.}
\label{Fig5}
\end{figure}

When we consider using flying atoms to convert a defective atom array to a defect-free one, the main advantage is that these atoms can be directly sent to vacancies without interfering other optical tweezers or atoms en route. In Fig.~\ref{Fig5}(a), we performed a proof-of-principle experiment. We first trapped eight atoms on a defective 3-by-3 square lattice (of lattice constant $\Lambda=4.2$~$\mu$m) using static optical tweezers, where the center ($A$ site) was vacant.  Then we trapped another atom (denoted by $A'$) using a dynamic optical tweezer and the atom was sent to fill the $A$ site, passing through an atom ($B$) en route. The resulting atom array is a successful defect-free array as in Fig.~\ref{Fig5}(b). In this demonstration, the potential depths $U_c=1.94$~mK and $U_s=0.58$~mK of the dynamic and static optical tweezers are chosen to minimize the loss of atoms. In comparison, when the dynamic optical tweezer, which carries the atom $A$, is guided, i.e., not turned off, while passing the atom $B$, the resulting array is another kind of defective atom array, as shown in Fig.~\ref{Fig5}(c), as the dynamic optical tweezer kicks out the atom at $B$. The defect-free array forming probabilities are compared between the usages of flying atom and guided atom in Fig.~\ref{Fig5}(d). 

Now to discuss the speed-up advantage of flying atoms in making a defect-free atom array, over conventional atom-guiding methods, we first consider a vacant site at the center of a linear chain of $N$ atoms. If we attempt to fix this array using sequentially guiding optical tweezers, i.e., scooting the atoms one by one from adjacent sites, the time required to complete necessary moves are given by $t_{\rm g}= 2N \sqrt{\Lambda /a_{\rm max}}$, where we assume the first half of the travel distance $\Lambda$ of each guiding is used for acceleration and the other half is for deceleration. In comparison, if we directly throw the boundary atom and catch it at the center, we can use the full distance $\Lambda$ for both acceleration and deceleration, so the time required for the vacancy filling is reduced by a factor of two, i.e., $t_{\rm f}\approx t_{\rm g}/2$. For more general cases of half-filled random array in one, two, or three dimensional space, we can resort to Monte Carlo simulation. Numerically estimated the scalings of the rearrangement times of the flying atom methods are given to be $t_{\rm f}^{1D}=O(N^{2})$, $t_{\rm f}^{2D}=O(N^{3/2})$, and $t_{\rm f}^{3D}=O(N^{4/3})$, for the one, two, and three dimensional cases, respectively. In all these cases, like the upper example, we get always $t_{\rm f}<t_{\rm g}$, so the flying-atom methods are always faster than the sequential guiding methods. 

We can also compare the flying-atom method with the simultaneous atom guiding by holographic optical tweezers. Then, the moving speed of holographic optical tweezers is given by $v_0=d f_{\rm p }$, which is limited by the phase refresh (about 30 frames per second) rate $f_{\rm p}$ of the holographic device and the trap width $d$ of the optical tweezers, because the trap movement induced by successive holograms must be smaller than $d$. So, the required time for a defect-free atom array by holographic optical tweezers is given by $t_{\rm H}\approx l/d f_{\rm p}$ of no $N$ scaling; however, with typical technical values~\cite{HyosubOE2019_imaging} of $f_{\rm p}=40$ Hz, $d=1$~$\mu$m, and $l=3$~$\mu$m, we can estimate $t_{\rm F}<t_{\rm H}$ for $N\lesssim 3000$.  In numerical estimation of random arrays, holographic rearrangement times are estimated to be $t_{\rm h}^{1D}=O(N)$, $t_{\rm h}^{2D}=O(N^{1/2})$, and $t_{\rm h}^{3D}=O(N^{1/3})$. So, we expect that the flying atom methods can be faster than the holographic simultaneous guidings for $N\lesssim 4000$ for 2D and $\lesssim 3000$ for 3D. 

In the context of neutral-atom acceleration, it may be worthwhile to compare the optical tweezer acceleration with Zeeman slowers. Typical Zeeman slowers are of $a_{\rm{max}} \approx 5\times 10^4$~$\rm{m/s^2}$, limited by atomic properties~\cite{zeeman_slower_limits}, while our current optical tweezer decelerator already achieves $a_{\rm{max}} = 2\times10^5$~$\rm{m/s^2}$. In a practical design, we could achieve a $\times 10$ advantage. In addition, as the typical atom lifetime inside an optical tweezer is $\tau_{\rm{atom}}= 40 s = 10^5 t_1$~\cite{KimPRXq2020_3DIsing}, if a technique to increase the deceleration length would be resolved, increasing the catching velocity from $v_0 =a_{\rm{typ}}t_1 \approx 1$~m/s to the above $v_0=a_{\rm{typ}}\tau_{\rm{atom}} > 400$~m/s would make an optical tweezer decelerator outperform Zeeman slowers. Also the flying atom method can create Rydberg atoms directly from flying atoms, which enables new kinds of Rydberg atom experiments such as single-atom collisions experiments, similarly as in charged particle cases~\cite{atom_ionpairs_collision}, and a new annealing method, like annealing with changing parameters of van der Waals interactions of Rydberg atoms which doesn't affect by laser phase noise~\cite{rydberg_repulsive_force}. 

In quantum information processing, most of the present quantum computing architectures are static, so their qubit-to-qubit interactions are local. However, flying atoms could build a dynamic quantum architecture. In quantum gate-based computation, for example, atomic qubits could be sequentially dragged to predefined one- or two-qubit gate operation zones so that a quantum circuit can be processed in different ways. Flying atoms can travel, while maintaining their quantum information, from one qubit system to another remote qubit system, potentially scaling the total qubit system. As a final remark, flying atoms are flying quantum memories, which combines the concepts in quantum communication and quantum computing, potentially paving a new route of quantum technologies.

\section{Conclusion}
In summary, we have presented single-atom throw-and-catch experiments, in which optical tweezers are used as an atom accelerator and decelerator to demonstrate flying atoms that can even penetrate other optical tweezers. These flying atoms are of a practical advantage in defect-free atom array formation in Rydberg-atom quantum computing and also suggest promising future applications in quantum information processing, such as flying quantum memories and flying-atom-based quantum computing, and fundamental studies of single-particle collisions.

\begin{acknowledgements} \noindent
This research was supported by Samsung Science and Technology Foundation (SSTF-BA1301-52).
\end{acknowledgements}

\end{document}